\documentclass[12pt]{article}
 \def\lddots{\mathinner{\mkern1mu\raise1pt\hbox{.}\mkern2mu
    \raise4pt\hbox{.}\mkern2mu\raise7pt\vbox{\kern7pt\hbox{.}}\mkern1mu}}
\makeatletter
\def\numberbysection{\@addtoreset{equation}{section}
 	\def\theequation{\thesection.\arabic{equation}}}
\makeatother

\numberbysection

\newcommand{\be}{\begin{eqnarray}}
\newcommand{\ee}{\end{eqnarray}}
\newcommand{\non}{\nonumber}
\newcommand{\n}{\ensuremath{\mathcal{N}}}
\newcommand{\tr}{\mathop{\rm tr}\nolimits}

\begin{document}

\begin{titlepage}
\strut\hfill DTP/00/19
\vspace{.5in}
\begin{center}

\LARGE Fusion and Analytical Bethe Ansatz for the $A_{\n-1}^{(1)}$ Open Spin Chain\\[1.0in]
\large Anastasia Doikou\\[0.8in]
\large Department of Mathematical Sciences, University of Durham,\\[0.2in]
\large South Road, DH1 3LE, Durham, United Kingdom\\

\end{center}

\vspace{.5in}

\begin{abstract}

We generalise the fusion procedure for the $A_{\n-1}^{(1)}$ open spin chain ($\n > 2$) and we show that the transfer matrix satisfies a crossing property. We use these results to solve the $A_{\n-1}^{(1)}$ open spin chain with $U_{q}\left(SU(\n)\right)$ symmetry by means of the analytical Bethe ansatz method. Our results coincide with the known ones obtained via the nested Bethe ansatz. \
\end{abstract}

\end{titlepage}

\section{Introduction}

We consider the $A_{\n - 1}^{(1)}$ open spin chain ($\n > 2$) with N sites \cite{bazhanov}, \cite{jimbo}. This is an integrable system which has been solved by means of the nested Bethe ansatz \cite{DVGR3}, \cite{DVGR2}. Also, this model, in the critical regime, can be thought as the lattice analogue of a certain two dimensional field theory i.e., the $A_{\n -1}^{(1)}$ affine Toda field theory \cite{zinn}. We focus here in the special case that the chain has $U_{q} \left(SU(\n) \right)$ symmetry \cite{mn/mpla}, \cite{mn/addendum}, and we solve it using a simpler method compared to nesting, namely, the analytical Bethe ansatz \cite{reshe}, \cite{mn/anal}. 

The analytical Bethe ansatz has been used for models with crossing symmetry e.g. $A_{1}^{(1)}$, $A_{2}^{(2)}$, $A_{2n}^{(2)}$, (see \cite{mn/anal}, \cite{amn/spectrum}). This is the first time that this method has been used for a model without crossing symmetry i.e., the $A_{\n - 1}^{(1)}$ spin chain. Although this chain does not have such symmetry, the corresponding  $R$ matrix satisfies a crossing property \cite{ogiev}-\cite{delius}, and consequently one can show that the transfer matrix satisfies an analogous property. One can also generalise the results of \cite{karowski}, \cite{mn/fusion} and derive a fusion procedure for the corresponding open chain transfer matrix. The crossing property, the fusion of the transfer matrix, and the quantum group symmetry play an essential role in the derivation of the analytical Bethe ansatz.

The outline of the paper is as follows: In section 2 we describe the model and we derive the crossing property for the $R$ matrix and the transfer matrix. In the next section  we deduce the fusion procedure for the open chain transfer matrix. In section 4 we derive the asymptotic behaviour of the transfer matrix and together with the results of the previous sections, periodicity and analyticity we find the spectrum of the transfer matrix and the Bethe ansatz equations. We illustrate the method using the $A_{2}^{(1)}$ chain, but the results can nevertheless be generalised for any $\n$. Finally, in the last section we give a brief discussion about the results of this work.

\section{The model}
There are two basic building blocks for constructing open spin chains:
\begin{enumerate}

\item The $R$ matrix, which is a solution of the Yang-Baxter equation
\be
R_{12}(\lambda_{1} - \lambda_{2})\ R_{1 3}(\lambda_{1})\ R_{23}(\lambda_{2}) = R_{23}(\lambda_{2})\ R_{1 3}(\lambda_{1})\ R_{1 2}(\lambda_{1} - \lambda_{2})\
\label{YBE}
\ee
(see, e.g., \cite{QISM}). We assume that the $R$ matrix has the unitarity property
\be
R_{12}(\lambda)R_{21}(-\lambda)= \zeta(\lambda) \,,
\label{property1}
\ee
where $R_{21}(\lambda) = {\cal P}_{12} R_{12}(\lambda) {\cal P}_{12} = R_{12}(\lambda)^{t_{1}t_{2}}$, $t$ denotes transpose, and $\cal P$ is the permutation matrix, and also the property \cite{RSTS}
\be
R_{12}(\lambda)^{t_{1}} M_{1} R_{12}(-\lambda - 2 \rho)^{t_{2}} M_{1}^{-1} = \zeta'(\lambda) \,,
\label{property2}
\ee
with  $M^{t} = M$,
\be
\left [M_{1} M_{2}\,, R_{12}(\lambda) \right ] = 0\,.
\label{property3}
\ee
and
\be
\zeta(\lambda)  = \sinh \mu (\lambda+i)  \sinh \mu (-\lambda+i)\,, \qquad
\zeta'(\lambda) = \sinh \mu (\lambda + \rho) \sinh \mu (-\lambda + \rho)\,.
\ee

For the purposes of this work we are going to need also the $R$ matrix that involves different representations of $U_q(SU(\n))$ \cite{VEWO}, \cite{abad}, in particular, $\n$ and $\bar \n$. This matrix is given by crossing
\be
R_{\bar 12}(\lambda) = V_{1}\ R_{12}(-\lambda - \rho)^{t_{2}}\ V_{1} = V_{2}^{t_{2}}\ R_{12}(-\lambda - \rho)^{t_{1}}\ V_{2}^{t_{2}} \,,
\label{prop4}
\ee
where $V^{2}=1$ and $M=V^{t}V$, for the $A_{1}^{(1)}$ case $R_{\bar1 2}(\lambda)= R_{12}(\lambda)$. $R_{\bar 12}(\lambda)$ also satisfies the unitarity property,
\be
R_{\bar 12}(\lambda)R_{2\bar 1}(-\lambda)= \zeta'(\lambda) \,,
\label{propery1p}
\ee
this equation is equivalent to (\ref{property2}), with $R_{2\bar 1}(\lambda) = R_{\bar 12}(\lambda)^{t_{1}t_{2}}$. Moreover 
\be
R_{\bar 12}(\lambda)^{t_{1}} M_{1} R_{\bar 12}(-\lambda - 2 \rho)^{t_{2}} M_{1}^{-1} = \zeta(\lambda),\
\label{property2p}
\ee
which is equivalent to (\ref{property1}).
$R_{\bar1 2}(\lambda)$ is also a solution of  the Yang-Baxter equation
\be
R_{\bar 1 2}(\lambda_{1} - \lambda_{2})\ R_{\bar1 3}(\lambda_{1})\ R_{23}(\lambda_{2}) = R_{23}(\lambda_{2})\ R_{\bar1 3}(\lambda_{1})\ R_{\bar1 2}(\lambda_{1} - \lambda_{2})\,.
\label{YBE2}
\ee
\item The matrices $K^{-}$, and $K^{+}$ which are solutions
of the boundary Yang-Baxter equation \cite{cherednik}
\be
R_{12}(\lambda_{1}-\lambda_{2})\ K_{1}^{-}(\lambda_{1})\
R_{21}(\lambda_{1}+\lambda_{2})\ K_{2}^{-}(\lambda_{2}) \non \\
= K_{2}^{-}(\lambda_{2})\ R_{12}(\lambda_{1}+\lambda_{2})\
K_{1}^{-}(\lambda_{1})\ R_{21}(\lambda_{1}-\lambda_{2}) \,,
\label{boundaryYB1}
\ee
and,
\be
R_{12}(-\lambda_{1}+\lambda_{2})\ K_{1}^{+}(\lambda_{1})^{t_{1}}\ M_{1}^{-1}\
R_{21}(-\lambda_{1}-\lambda_{2}-2\rho)\ M_{1}\ K_{2}^{+}(\lambda_{2})^{t_{2}} \non \\
= K_{2}^{+}(\lambda_{2})^{t_{2}}\ M_{1}\ R_{12}(-\lambda_{1}-\lambda_{2}-2\rho)\
M_{1}^{-1}\ K_{1}^{+}(\lambda_{1})^{t_{1}}\ R_{21}(-\lambda_{1}+\lambda_{2}) \,,
\label{boundaryYB2}
\ee
there exist an automorphism between $K^{-}$ and $K^{+}$ i.e.,
\be
K^{+}(\lambda) = M K^{-}(-\lambda-\rho)^{t}\,.
\ee
For the following we are going to need a reflection equation that involves
$R_{\bar1 2}$ as well, in particular,
\be
R_{\bar1 2}(\lambda_{1}-\lambda_{2})\ K_{\bar1}^{-}(\lambda_{1})\
R_{2\bar1}(\lambda_{1}+\lambda_{2})\ K_{2}^{-}(\lambda_{2}) \non \\
= K_{2}^{-}(\lambda_{2})\ R_{\bar12}(\lambda_{1}+\lambda_{2})\
K_{\bar1}^{-}(\lambda_{1})\ R_{2\bar1}(\lambda_{1}-\lambda_{2}) \,,
\label{bYBc}
\ee
and,
\be
R_{\bar12}(-\lambda_{1}+\lambda_{2})\ K_{\bar1}^{+}(\lambda_{1})^{t_{1}}\ M_{1}^{-1}\
R_{2\bar1}(-\lambda_{1}-\lambda_{2}-2\rho)\ M_{1}\ K_{2}(\lambda_{2})^{t_{2}} \non \\
= K_{2}^{+}(\lambda_{2})^{t_{2}}\ M_{1}\ R_{\bar12}(-\lambda_{1}-\lambda_{2}-2\rho)\
M_{1}^{-1}\ K_{\bar1}^{+}(\lambda_{1})^{t_{1}}\ R_{2\bar1}(-\lambda_{1}+\lambda_{2}) \,.
\label{bYB2c}
\ee
In the scattering language if we think that  the $K_{i}$ matrix describes the scattering of a soliton with the boundary, then $K_{ \bar i}$ describes the scattering of an anti-soliton with the boundary. Subsequently $R_{\bar 1 2}$ describes the scattering of a soliton with an anti-soliton.

\end{enumerate}

The corresponding transfer matrix $t(\lambda)$ for an open chain of
$N$ spins is given by
\cite{sklyanin}, \cite{mn/nonsymmetric}
\be
t(\lambda) = \tr_{0}  K_{0}^{+}(\lambda)\
T_{0}(\lambda)\  K^{-}_{0}(\lambda)\ \hat T_{0}(\lambda)\,,
\label{transfer1}
\ee
where $\tr_{0}$ denotes trace over the ``auxiliary space'' 0,
$T_{0}(\lambda)$ is the monodromy matrix
\be
T_{0}(\lambda) = R_{0N}(\lambda) \cdots  R_{01}(\lambda) \,,
\label{monodromy}
\ee
and
$\hat T_{0}(\lambda)$ is given by
\be
\hat T_{0}(\lambda) = R_{10}(\lambda) \cdots  R_{N0}(\lambda) \,.
\label{hatmonodromy}
\ee
(We usually suppress the ``quantum-space'' subscripts
$1 \,, \ldots \,, N$.) Indeed, it can be shown that this transfer
matrix has the commutativity property
\be
\left[ t(\lambda)\,, t(\lambda') \right] = 0 \,.
\label{commutativity}
\ee

In this paper, we consider the case of the $A_{\n-1}^{(1)}$ $R$ matrix
\cite{devega}
\be
R_{12}(\lambda)_{j j \,, j j} &=& {\sinh \mu(\lambda + i)} \,, \non \\
R_{12}(\lambda)_{j k \,, j k} &=& {\sinh (\mu \lambda)} \,, \qquad j \ne k
\,, \non \\
R_{12}(\lambda)_{j k \,, k j} &=& {\sinh (i \mu )}e^{\mu \lambda
sign(j-k)} \,, \qquad j \ne k \,, \non \\
& & 1 \le j \,, k \le \n \,,
\label{Rmatrix}
\ee
which depends on the anisotropy parameter $\mu \ge 0$, and which becomes
$SU(\n)$ invariant for $\mu \rightarrow 0$. This $R$ matrix has the properties
(\ref{property1}) and (\ref{property2}), with \cite{DVGR0}
\be
M_{j k} = \delta_{j k} e^{-i \mu (\n - 2 j + 1) }\,, \qquad \
\rho = i\n/ 2 \,.
\label{M}
\ee

The corresponding open spin chain Hamiltonian $\cal H$ for $K^{-}(\lambda) = 1$ and $K^{+}(\lambda) = M$ is:
\be
{\cal H} = \sum_{n=1}^{N-1} {\cal H}_{n n+1}
 + {\tr_{0} M_{0} 
{\cal H}_{N 0}\over \tr M}\,, 
\ee
where the two-site Hamiltonian ${\cal H}_{j k}$ is given by
\be
{\cal H}_{j k} = {i \over 2} {\cal P}_{j k} {d\over d \lambda}
R_{j k}(\lambda) \Big\vert_{\lambda=0}  \,.
\ee
One can verify that the Hamiltonian is Hermitian.

We consider the case that $K^{-}(\lambda) = 1$ and $K^{+}(\lambda) = M$,
and so the transfer matrix is $U_{q}\left(
SU(\n) \right)$ invariant \cite{mn/mpla}, \cite{mn/addendum}.
Following \cite{mn/anal} we show that the transfer matrix satisfies a crossing property. To prove the crossing property we need (\ref{YBE}), (\ref{property2}) and the  following identity
\be
{\cal P}_{12}^{t_{2}}\ M_{2}\ R_{21}(\lambda)^{t_{1}} =  R_{21}(\lambda)^{t_{1}}\
M_{1}^{-1}\ {\cal P}_{12}^{t_{2}} \,.
\label{bound}
\ee
It is important to mention that in order to show (\ref{bound}) we considered the ``unusual''
reflection equation (\ref{bYBc}) for $\lambda_{1} - \lambda_{2} = -\rho$.
Then the crossing property for the transfer matrix is given by
\be
t(\lambda) = \bar t(-\lambda -\rho) \,,
\label{cross}
\ee
where
\be
\bar t(\lambda) = \tr_{0} M_{0}\ 
T_{\bar0}(\lambda)  \hat T_{\bar0}(\lambda)\,,
\label{transfer12} 
\ee
and
\be
T_{\bar0}(\lambda)      &=& R_{\bar0 N}(\lambda) \cdots  R_{\bar0 1}(\lambda)\,, \non\\
\hat T_{\bar0}(\lambda) &=& R_{1 \bar0}(\lambda) \cdots  R_{N \bar0}(\lambda),\
\label{hatmonodromyb}
\ee
the proof of equation (\ref{cross}) follows exactly the proof in \cite{mn/anal}. The only difference is that in this case $\bar t(\lambda)$ is involved as well because of (\ref{prop4}). The transfer matrix $\bar t(\lambda)$ satisfies the commutativity property 
\be
\left[ \bar t(\lambda)\,, \bar t(\lambda') \right] = 0 \,.
\label{commutativityb}
\ee
Relation (\ref{cross}) is one of the basic results of this paper and it plays an essential role in the derivation of the transfer matrix eigenvalues.
The ``new'' transfer matrix (see also e.g., \cite{VEWO}) leads apparently to a non local Hamiltonian, however, this is not a problem since ${\bar t(\lambda)}$ has  an auxiliary character in our calculations as we are going to see later.

\section{Fusion}

The fusion procedure for spin chains with crossing symmetry is known 
\cite{mn/fusion}. We generalise this procedure for the case that the $R$ matrix does not have crossing symmetry. From now on the indices 1 and 2 refer to the auxiliary space. We consider the equation (\ref{prop4}) for $\lambda = -\rho$. Then
the $R_{\bar1 2}(\lambda)$ degenerates to a projector onto an one dimensional subspace
\be
P_{\bar 1 2}^{-} = {1 \over \n}\ V_{1}\ {\cal P}_{12}^{t_2}\ V_{1}\,, 
\label{proj}
\ee
also
\be
P_{\bar1 2}^{+} = 1 - P_{\bar1 2}^{-}\
\label{proj+}
\ee
is a projector.
We consider the Yang-Baxter equation (\ref{YBE2}) for $\lambda_{1}- \lambda_{2} = -\rho$, then the fused $R$ matrix is given by
\be
R_{<\bar1 2>3}(\lambda) &=& P_{\bar1 2}^{+}\ R_{\bar1 3}(\lambda)\ R_{23}(\lambda + \rho)\ P_{\bar1 2}^{+}\,,\non\\
R_{<2 \bar1 >3}(\lambda) &=& P_{2 \bar1}^{+}\ R_{23}(\lambda)\ R_{\bar1 3}(\lambda + \rho)\ P_{2 \bar1}^{+} \,,
\label{fusion12}
\ee
also, one finds
\be
R_{3<\bar1 2>}(\lambda) &=& P_{\bar1 2}^{+}\ R_{32 }(\lambda - \rho)\ R_{3 \bar1}(\lambda)\ P_{\bar1 2}^{+}\,,\non\\
R_{3<2 \bar1 >}(\lambda) &=& P_{2 \bar1}^{+}\ R_{3 \bar1}(\lambda - \rho)\ R_{32}(\lambda)\ P_{2 \bar1}^{+} \,,
\label{fusion12p}
\ee
similarly, one can fuse the spaces 1 and $\bar 2$. The fused $R$ matrices obey the general Yang-Baxter equation
\be
R_{j_{1}j_{2}}(\lambda_{1})\ R_{j_{1}j_{3}}(\lambda_{1} + \lambda_{2})\ R_{j_{2}j_{3}}(\lambda_{2})
= R_{j_{2}j_{3}}(\lambda_{2})\ R_{j_{1}j_{3}}(\lambda_{1} + \lambda_{2})\ R_{j_{1}j_{2}}(\lambda_{1}) \,.
\label{YBg}
\ee

Consider the reflection equation (\ref{bYBc}) for $\lambda_{1}-\lambda_{2}=-\rho$, then the fused $K$ matrices are given by
\be
& &K^{-}_{<\bar1 2>}(\lambda) = P_{\bar1 2}^{+}\ K^{-}_{\bar1}(\lambda)\ R_{2 \bar1}(2 \lambda + \rho)\ K^{-}_{2}(\lambda+\rho)\ P_{2 \bar1}^{+} \,, \non \\
& &K_{<\bar1 2>}^{+}(\lambda)^{t_{12}} = P_{2 \bar1 }^{+}\ K_{\bar1}^{+}(\lambda)^{t_{1}}\ M_{2}\ R_{2 \bar1}(-2 \lambda -3 \rho)\ M_{2}^{-1} K_{2}^{+}(\lambda+\rho)^{t_{2}}\ P_{ \bar1 2 }^{+} \,.
\ee
The above $K$ matrices obey the reflection equations
\be
R_{3<\bar12>}(\lambda_{1}-\lambda_{2})\ K_{3}^{-}(\lambda_{1})\
R_{<\bar1 2>3}(\lambda_{1}+\lambda_{2})\ K_{<\bar12>}^{-}(\lambda_{2}) \non \\
= K_{<\bar1 2>}^{-}(\lambda_{2})\ R_{3<2\bar1>}(\lambda_{1}+\lambda_{2}+\rho)\
K_{3}^{-}(\lambda_{1})\ R_{<2\bar1>3}(\lambda_{1}-\lambda_{2}-\rho) \,,
\label{boundaryYBf}
\ee
and,
\be
R_{<\bar12>3}(-\lambda_{1}+\lambda_{2})^{t_{123}}\ K_{3}^{+}(\lambda_{1})^{t_{3}}\ M_{3}^{-1}\ R_{3<\bar12>}(-\lambda_{1}-\lambda_{2}-2\rho)^{t_{123}}\ M_{3}\ K_{<\bar12>}^{+}(\lambda_{2})^{t_{12}} 
\non \\
= K_{<\bar12>}^{+}(\lambda_{2})^{t_{12}}\ M_{3}\ R_{<2\bar1>3}(-\lambda_{1}-\lambda_{2}-3\rho)^{t_{123}}\ M_{3}^{-1}\ 
K_{3}^{+}(\lambda_{1})^{t_{3}}\ R_{3<2\bar1>}(-\lambda_{1}+\lambda_{2}+\rho)^{t_{123}} \,,
\label{boundaryYB2f}
\ee
analogously we obtain the $K_{<1 \bar 2>}(\lambda)$ matrices.

Having fused the $R$ and $K$ matrices we can show that the corresponding fused transfer matrix is (for a detailed computation see e.g., \cite{mn/fusion})
\be
\hat t(\lambda) = \zeta(2\lambda+2\rho)\ \bar t(\lambda)\ t(\lambda + \rho) - \Delta[K^{+}(\lambda)]\ \delta[T(\lambda)]\ \Delta[K^{-}(\lambda)]\ \delta[\hat T(\lambda)]\,, 
\label{fusiont}
\ee
notice that $\bar t(\lambda)$ appears in the last equation as well as in (\ref{cross}). 
In this case we fuse the spaces $\bar 12$, and the quantum determinants are 
\cite{sklyanin}, \cite{mn/fusion},
\be
& &\delta[ T(\lambda)]     = tr_{12}[P^{-}_{\bar12}\ T_{\bar1}(\lambda)\ \hat T_{2}(\lambda+\rho)]\,, \non\\
& &\delta[\hat T(\lambda)] = tr_{12}[P^{-}_{2\bar1}\ T_{\bar1}(\lambda)\ \hat T_{2}(\lambda+\rho)]\,, \non\\
& &\Delta[K^{-}(\lambda)]  = tr_{12}[P^{-}_{\bar1 2}\ K^{-}_{1}(\lambda)\ R_{2 \bar1}(2\lambda + \rho)\ K^{-}_{2}(\lambda+\rho)\ V_{1}\ V_{2}] \,, \non\\
& &\Delta[K^{+}(\lambda)]  = tr_{12}[P^{-}_{\bar1 2}\ V_{1}\ V_{2}\ K^{+}_{2}(\lambda+ \rho)\ M_{2}^{-1}\ R_{\bar1 2}(-2\lambda -3 \rho)\ M_{2}\ K^{+}_{1}(\lambda)] \,. 
\label{det}
\ee
Similar relations to (\ref{fusiont}) and (\ref{det}) we can derive for the $1 \bar 2$ fusion.
Using unitarity and the crossing property we prove the identities
\be
P^{-}_{\bar1 2}\ R_{\bar1m}(\lambda)\ R_{2m}(\lambda+\rho)\ P^{-}_{\bar1 2} &=&
 \zeta(\lambda+\rho) P^{-}_{\bar1 2}\,,\non\\
P^{-}_{1\bar2}\ R_{1m}(\lambda)\ R_{\bar2m}(\lambda+\rho)\ P^{-}_{1\bar2}   &=&
 \zeta'(\lambda+\rho) P^{-}_{1\bar2}\,, \non\\
m=1,2,\ldots,N
\label{ident}
\ee
and by computing the quantum determinants explicitly we find
\be
\delta[T(\lambda)] = \delta[\hat T(\lambda)] = \zeta(\lambda+\rho)^{N}\ or\ \zeta'(\lambda+\rho)^{N}\,,
\label{Tdet}
\ee
depending on the spaces we fuse, i.e., $\bar12$ or $1\bar2$ respectively.
For the special case $K^{-}(\lambda)=1$
and $K^{+}(\lambda)= M$, these are solutions of the reflection equations (\ref{bYBc}) and (\ref{bYB2c}) correspondingly, one can show from (\ref{det}) that,
\be
\Delta[K^{-}(\lambda)] = g(2\lambda + \rho)\,,\qquad\
\Delta[K^{+}(\lambda)] = g(-2\lambda -3\rho)\,,
\ee
where,
\be
g(\lambda) = \sinh \mu (-\lambda + \rho).\
\ee

\section{Analytical Bethe Ansatz}

We focus here in the simplest case, namely, the $A_{2}^{(1)}$ open chain.
 The asymptotic behaviour of the $R$ matrix for $\lambda \rightarrow -\infty$ follows from (\ref{Rmatrix}) 
\be
R_{0k} & \sim & - {1 \over 2}\ e^{-\mu\lambda} \left(
           \begin{array}{ccc}
            e^{-i \mu S_{1,k}} &pJ^{-}_{1,k}       &pJ^{-}_{3,k}  \\
            0                  &e^{-i \mu S_{2,k}} &pJ^{-}_{2,k}   \\                      0                  &0                  &e^{-i \mu S_{3,k}}\\ 
                                                                         \end{array}\right)\,,\non\\
R_{k0} & \sim & - {1 \over 2}\ e^{-\mu\lambda} \left
           (\begin{array}{ccc}
            e^{-i \mu S_{1,k}} &0                  &0                 \\
            pJ^{+}_{1,k}       &e^{-i \mu S_{2,k}} &0                 \\                   pJ^{+}_{3,k}       &pJ^{+}_{2,k}       &e^{-i \mu S_{3,k}} 
                                                                         \end{array} \right),\
\label{asympt}
\ee 
where $p = 2 \sinh(-i \mu)$, and the matrix elements are:
\be
S_{i}     &=& e_{i,i}\,, \qquad i=1,2,3\,, \non\\
J^{+}_{i} &=& e_{i,i+1}\,, \qquad
J^{-}_{i} = e_{i+1,i}\,, \qquad
 i=1,2\,, \non\\
J^{+}_{3} &=& e_{1,3}\,,\qquad
J^{-}_{3} = e_{3,1}\,,
\ee 
with,
\be
(e_{i,j})_{kl} = \delta_{ik} \delta_{jl}\,.
\label{e}
\ee
The leading asymptotic behaviour of the monodromy matrix is given by
\be
T_{0}(\lambda)      & \sim & (-{1 \over2})^{N}\ e^{-\mu \lambda N} \left
       (\begin{array}{ccc}
        e^{-i \mu {\cal S}_{1}}  &p{\cal J}_{1}^{-}       &p{\cal J}_{3}^{-}\\ 
        0                        &e^{-i \mu {\cal S}_{2}} &p{\cal J}_{2}^{-} \\        0                        &0                       &e^{-i \mu {\cal S}_{3}}\\ 
                                                                         \end{array}\right)\,,\non\\
\hat T_{0}(\lambda)  & \sim &  (-{1 \over2})^{N}\ e^{-\mu \lambda N} \left
       (\begin{array}{ccc}
        e^{-i \mu {\cal S}_{1}}  &0                       &0           \\
        p{\cal J}_{1}^{+}        &e^{-i \mu {\cal S}_{2}} &0                 \\        p{\cal J}_{3}^{+}        &p{\cal J}_{2}^{+}       &e^{-i \mu {\cal S}_{3}}\\ 
                                                                         \end{array}\right) \,,
\label{asymptT}
\ee
where,
\be
{\cal S}_{i}      &=& \sum_{k=1}^{N}S_{i,k}\, \qquad i=1,2,3\,,\non\\
{\cal J}_{1}^{\pm}&=& \sum_{k=1}^{N}e^{-i \mu S_{1,N}} \ldots e^{-i \mu S_{1,k+1}} J_{1,k}^{\pm} e^{-i \mu S_{2,k-1}} \ldots e^{-i \mu S_{2,1}}\,,\non\\
{\cal J}_{2}^{\pm}&=& \sum_{k=1}^{N}e^{-i \mu S_{2,N}} \ldots e^{-i \mu S_{2,k+1}} J_{2,k}^{\pm} e^{-i \mu S_{3,k-1}} \ldots e^{-i \mu S_{3,1}}\,, \non\\
{\cal J}_{3}^{\pm}&=& \sum_{k=1}^{N}(e^{-i \mu S_{1,N}} \ldots e^{-i \mu S_{1, k+1}}J_{3,k}^{\pm} + p {\cal J}_{1,k+1}^{\pm} J_{2,k}^{\pm}) e^{-i \mu S_{3,k-1}} \ldots e^{-i \mu S_{3,1}} \,. 
\label{quant}
\ee
Then from equations (\ref{transfer1}) for $K^{-}(\lambda) = 1$ and $K^{+}(\lambda) = M$, (\ref{M}), and (\ref{asymptT}) we conclude that the leading asymptotic behaviour of the transfer matrix has the following form
\be
t(\lambda) & \sim & ({1 \over2})^{2N}e^{-2 \mu \lambda N}(e^{ -2i \mu-2i \mu {\cal S}_{1}} + e^{-2i\mu}p^{2}{\cal J}_{1}^{-}{\cal J}_{1}^{+} + e^{ -2i \mu} p^{2} {\cal J}_{3}^{-}{\cal J}_{3}^{+}\ \non\\ 
           &+& e^{2i\mu - 2i\mu {\cal S}_{3}} + p^{2} {\cal J}_{2}^{-} {\cal J}_{2}^{+} +e^{-2 i\mu {\cal S}_{2}})\,.
\label{aseig}
\ee
We introduce the operators $M_{1}$ and $M_{2}$ 
\be
{\cal S}_{1} = N - M_{1}\,,\qquad {\cal S}_{2} = M_{1} - M_{2}\,,\qquad {\cal S}_{3} = M_{2}\,,
\label{Mo}
\ee
we consider simultaneous eigenstates of $M_{i}$ and the transfer matrix i.e.,
\be
M_{i}|\Lambda^{(m)} \rangle \ = m_{i} |\Lambda^{(m)} \rangle \,, \qquad t(\lambda)|\Lambda^{(m)} \rangle\ = \Lambda^{(m)}(\lambda) |\Lambda^{(m)} \rangle\,,
\label{eigen}
\ee
(to simplify our notation we write $(m)$ instead of $(m_{1}m_{2})$).
We choose these states to be annihilated by ${\cal J}_{i}^{+}$
\be
{\cal J}_{i}^{+}|\Lambda^{(m)} \rangle = 0\,.
\label{hi}
\ee
We conclude, from (\ref{aseig}), (\ref{eigen}), and (\ref{hi}) that the asymptotic behaviour of the corresponding eigenvalue is given by
\be
\Lambda^{(m)}(\lambda) \sim ({1 \over2})^{2N} e^{-2 \mu \lambda N}(e^{-2i\mu (1+N-m_{1})} + e^{2i\mu (1-m_{2})} + e^{-2i\mu (m_{1}-m_{2})})\,.
\label{asyme}
\ee
In order to determine the asymptotic behaviour of $\bar t(\lambda)$, we also need the asymptotic behaviour of $R_{\bar0k}(\lambda) (R_{k \bar0}(\lambda))$ for $\lambda \rightarrow -\infty$
\be
R_{\bar 0k} & \sim & {1 \over 2}\ e^{-\mu\lambda - {3i \mu \over 2}} \left(
           \begin{array}{ccc}
            e^{i \mu S_{3,k}} &qpJ^{-}_{2,k}      &-q^{2}pJ^{-}_{3,k}  \\
            0                 &e^{i \mu S_{2,k}}  &qpJ^{-}_{1,k}   \\                      0                 &0                  &e^{i \mu S_{1,k}}\\ 
                                                                         \end{array}\right)\,,\non\\
R_{k \bar0} & \sim & {1 \over 2}\ e^{-\mu\lambda - {3i \mu \over 2}} \left
           (\begin{array}{ccc}
            e^{i \mu S_{3,k}}  &0                 &0                 \\
            qpJ^{+}_{2,k}      &e^{i \mu S_{2,k}} &0                 \\                    -q^{2}pJ^{+}_{3,k} &qpJ^{+}_{1,k}     &e^{i \mu S_{1,k}} 
                                                                         \end{array} \right),\
\label{asympt2}
\ee
we define $R_{\bar0k}(\lambda)$ from (\ref{prop4})  using
\be
V  = \left(
       \begin{array}{ccc}
                &   &q  \\
                &-1 &     \\          
         q^{-1} &   &  \\ 
                                                                         \end{array}\right)\,
\ee
where $q= e^{i \mu}$, and the asymptotic behaviour of $\bar t(\lambda)$ is given by
\be
\bar t(\lambda) & \sim & ({1 \over2})^{2N}e^{-2 \mu \lambda N - 3i \mu N}(e^{ 2i \mu + 2i \mu {\cal S}_{1}} + e^{2i\mu} p^{2} \bar {\cal J}_{1}^{-} \bar {\cal J}_{1}^{+} + e^{2i\mu} p^{2}\bar {\cal J}_{3}^{-}\bar {\cal J}_{3}^{+}\ \non\\
                &+& e^{-2i\mu + 2i\mu {\cal S}_{3}} + p^{2}\bar  {\cal J}_{2}^{-}\bar {\cal J}_{2}^{+} +e^{2 i\mu {\cal S}_{2}}) \,,
\label{aseigp}
\ee
with
\be
\bar {\cal J}_{1}^{\pm}&=& \sum_{k=1}^{N}e^{i \mu S_{2,N}} \ldots e^{i \mu S_{2,k+1}} J_{1,k}^{\pm} e^{i \mu S_{1,k-1}} \ldots e^{i \mu S_{1,1}}\,,\non\\
\bar {\cal J}_{2}^{\pm}&=& \sum_{k=1}^{N}e^{i \mu S_{3,N}} \ldots e^{i \mu S_{3,k+1}} J_{2,k}^{\pm} e^{i \mu S_{2,k-1}} \ldots e^{i \mu S_{2,1}}\,, \non\\
\bar {\cal J}_{3}^{\pm}&=& \sum_{k=1}^{N}(e^{i \mu S_{3,N}} \ldots e^{i \mu S_{3, k+1}}J_{3,k}^{\pm} - p\bar {\cal J}_{2,k+1}^{\pm} J_{1,k}^{\pm}) e^{i \mu S_{1,k-1}} \ldots e^{i \mu S_{1,1}} \,.
\label{quantp}
\ee
The $|\Lambda^{(m)} \rangle$ states are also annihilated by $\bar {\cal J}_{i}^{+}$ i.e.,
\be
\bar {\cal J}_{i}^{+}|\Lambda^{(m)} \rangle = 0\,,
\ee
where the corresponding eigenvalue of $\bar t(\lambda)$ is
\be
\bar \Lambda^{(m)}(\lambda) \sim ({1 \over2})^{2N} e^{-2\lambda \mu N - 3i\mu N}(e^{2i\mu (1+N-m_{1})} + e^{-2i\mu (1-m_{2})} + e^{2i\mu (m_{1}-m_{2})})\,.
\label{asyme2}
\ee 

We consider the state with all ``spins'' up i.e.,
\be
|\Lambda^{(0)} \rangle =  \bigotimes_{k=1}^{N} |+ \rangle_{(k)}\,,
\label{state}
\ee
 this is annihilated by ${\cal J}_{i}^{+}$ and $\bar {\cal J}_{i}^{+}$, where (we suppress the $(k)$ index)
\be
|+ \rangle = \left (\begin{array}{c}
                     1 \\
                     0  \\ 
                     0  \\
                       \end{array} \right)\,.
\label{col}
\ee
We assume this is an eigenstate of the transfer matrix and it is also an eigenstate of $\bar t(\lambda)$. The action of the $R$ matrix on the $|+ \rangle$ state gives lower and upper triangular matrices
 i.e.,
\be
\langle +|R_{0k}(\lambda) = \langle +  | \left(
           \begin{array}{ccc}
            A_{k}   &0       &0  \\
            C_{1,k} &D_{1,k} &0        \\   
            C_{2,k} &D_{3,k} &D_{4,k}  \\ 
                                         \end{array}\right)\,,\qquad
R_{k0}(\lambda)|+ \rangle = \left
           (\begin{array}{ccc}
            A_{k} &B_{1,k} &B_{2,k}     \\
            0     &D_{1,k} &D_{2,k}          \\                                            0     &0       &D_{4,k}    \\
                                         \end{array} \right) |+ \rangle \,,
\label{matrices}
\ee 
where the matrices $A$, $B_{i}$, $C_{i}$, $D_{i}$ act on the quantum space and they are determined by the form of the $R$ matrix (\ref{Rmatrix}).
Then the action of the transfer matrix on the pseudo-vacuum is,
\be
\langle \Lambda^{(0)}|T_{0}(\lambda) = \langle \Lambda^{(0)}  | \left(
           \begin{array}{ccc}
           {\cal A}     &0              &0  \\
           {\cal C}_{1} &{\cal D}_{1}   &0        \\   
           {\cal C}_{2} &{\cal D}_{3}   &{\cal D}_{4}  \\ 
                                          \end{array}\right)\,, \qquad
\hat T_{0}(\lambda)|\Lambda^{(0)} \rangle = \left  
           (\begin{array}{ccc}
           {\cal A}     &{\cal B}_{1}   &{\cal B}_{2}      \\
            0           &{\cal D}_{1}   &{\cal D}_{2}           \\                         0           &0              &{\cal D}_{4}     \\
                                            \end{array} \right) |\Lambda^{(0)} \rangle \,,
\label{matricest}
\ee
and the matrix elements are given by
\be
{\cal A} = A_{N} \ldots A_{1}\,,\qquad {\cal D}_{1} = D_{1,N} \ldots D_{1,1}\,,\qquad
{\cal D}_{4} = D_{4,N} \ldots D_{4,1}\,,
\ee
\be
{\cal C}_{1} &=& \sum_{k=1}^{N} D_{1,N} \ldots D_{1,k+1} C_{1,k} A_{k-1} \ldots
A_{1}\,, \qquad {\cal C}_{0} = 0\,, \non\\
{\cal D}_{3} &=& \sum_{k=1}^{N} D_{4,N} \ldots D_{4,k+1} D_{3,k} D_{1,k-1} \ldots D_{1,1}\,, \qquad {\cal D}_{3,0}=0\,, \non\\
{\cal C}_{2} &=& \sum_{k=1}^{N}D_{4,N} \ldots D_{4,k+1}(C_{2,k}A_{k-1} \ldots
A_{1} + D_{3,k} {\cal C}_{1,k-1})\,.
\label{equ} 
\ee
Similarly, to find ${\cal B}_{1}$, ${\cal B}_{2}$ and ${\cal D}_{2}$ we replace $C_{1}$, $D_{3}$, $C_{2}$ and ${\cal C}_{1}$ with $B_{1}$, $D_{2}$, $B_{2}$ and ${\cal B}_{1}$ correspondingly.
The transfer matrix eigenvalue for the pseudo-vacuum state is
\be
\Lambda^{(0)}(\lambda) &=& \langle \Lambda^{(0)} |(e^{-2i\mu} {\cal A}^{2} + {\cal D}_{1}^{2} + e^{2i\mu} {\cal D}_{4}^{2} +{\cal C}_{1}{\cal B}_{1} + e^{2i\mu} {\cal C}_{2} {\cal B}_{2} + e^{2i\mu} {\cal D}_{3}{\cal D}_{2})|\Lambda^{(0)} \rangle\,, 
\ee
and after some tedious algebra we find
\be
\Lambda^{(0)}(\lambda)= f(\lambda)( a(\lambda)^{2N} {\sinh 2\mu(\lambda + i)
\over \sinh(2\mu \lambda)} + b(\lambda)^{2N}(1 + {\sinh \mu(2\lambda + i) \over \sinh \mu(2\lambda + 3i)}))\,.
\ee
 We make the assumption that a general eigenvalue has the form of a ``dressed'' pseudo-vacuum  eigenvalue i.e.,
\be
\Lambda^{(m)}(\lambda) = f(\lambda) (a(\lambda)^{2N} {\sinh 2\mu(\lambda + i) \over \sinh(2\mu \lambda)} A_{1}(\lambda) + b(\lambda)^{2N}( A_{2}(\lambda) +  {\sinh \mu (2\lambda + i) \over \sinh \mu(2\lambda + 3i)} A_{3}(\lambda)))\,.
\ee
Again, the action of $R_{\bar0k}(\lambda)$ on the $|+ \rangle$ state gives upper and lower triangular matrices (see (\ref{matrices})), so we find an analogous equation for the $\bar \Lambda^{(m)}(\lambda)$,
\be
\bar \Lambda^{(m)}(\lambda) = f(\lambda)(\bar a(\lambda)^{2N} {\sinh \mu(2\lambda + i) \over \sinh \mu(2\lambda + 3i)} \bar A_{1}(\lambda) + \bar b(\lambda)^{2N} ( \bar A_{2}(\lambda) + {\sinh 2\mu(\lambda + i)\over \sinh(2\mu \lambda)} \bar A_{3}(\lambda)))\,,
\ee
(we suppress the $(m)$ index from the ``dressing'' functions), where
\be
f(\lambda) = {\sinh(2 \mu \lambda) \sinh \mu(2 \lambda +3i) \over \sinh 2 \mu(\lambda +i) \sinh \mu(2 \lambda +i)}\,,
\ee 
\be
a(\lambda) &=& \sinh \mu(\lambda + i)\,, \qquad 
b(\lambda) = \sinh(\mu \lambda)\,,
\ee
and $\bar a(\lambda)$, $\bar b(\lambda)$ are $a(-\lambda-\rho)$, $b(-\lambda-\rho)$, respectively. It is obvious that $ \Lambda^{(0)}(\lambda) = \bar \Lambda^{(0)}(-\lambda - \rho)$.
We conclude from the asymptotic behaviour of the transfer matrix that
\be
A_{1}(\lambda) \rightarrow e^{2i \mu m_{1}}\,, \qquad 
A_{2}(\lambda) \rightarrow e^{2i\mu (m_{2} - m_{1})}\,, \qquad A_{3}(\lambda) \rightarrow e^{-2i \mu m_{2}}\,,
\ee
and
\be
\bar A_{1}(\lambda) \rightarrow e^{-2i \mu m_{1}}\,, \qquad 
\bar A_{2}(\lambda) \rightarrow e^{-2i \mu (m_{2} - m_{1})}\,, \qquad \bar A_{3}(\lambda) \rightarrow e^{2i \mu m_{2}}\,.
\ee
We substitute the eigenvalues to the fusion equations (\ref{fusiont}) and we obtain conditions involving $A_{1}(\lambda)$, $A_{3}(\lambda)$, and $\bar A_{1}(\lambda)$, $\bar A_{3}(\lambda)$. It is clear that the $\Lambda^{(0)}(\lambda)$ satisfies (\ref{fusiont}) a fact that further supports our assumption that $|\Lambda^{(0)} \rangle$ is an eigenstate of the transfer matrices, $t(\lambda)$ and $\bar t(\lambda)$. The fusion equations give
\be
A_{1}(\lambda + \rho) \bar A_{1}(\lambda) = 1\,, \qquad \bar A_{3}(\lambda + \rho)A_{3}(\lambda) = 1\,,
\ee
where notice that we obtain two conditions from (\ref{fusiont}), one from $\bar 1 2$ fusion and one from $1 \bar 2$, whereas e.g., for the $A_{1}^{(1)}$ case we obtain only one such condition. From the crossing property of the transfer matrix (\ref{cross}),
\be
A_{i}(-\lambda - \rho) = \bar A_{i}(\lambda)\,, \qquad  i=1,2,3\,.
\label{crosp}
\ee
Combining the last two conditions we find
\be
A_{1}(\lambda)A_{1}(-\lambda) = 1\,, \qquad \bar A_{3}(\lambda) \bar A_{3}(-\lambda) = 1\,.
\ee
Note that the previous equations mix the ``dressing'' functions of $\Lambda^{(m)}(\lambda)$ and $\bar \Lambda^{(m)}(\lambda)$, which is expected because of the form of equations (\ref{cross}) and (\ref{fusiont}). In the case of a model with crossing symmetry e.g., $A_{1}^{(1)}$, the two eigenvalues become degenerate. From the periodicity of the transfer matrix we obtain
\be
t(\lambda + {i \pi \over \mu}) = t(\lambda)\,,
\ee
we expect the eigenvalues to be periodic as well. We impose $A_{2}(\lambda)$ ($\bar A_{2}(\lambda)$) to have the same poles with $A_{1}(\lambda)$ and $A_{3}(\lambda)$ ($\bar A_{1}(\lambda)$ and $\bar A_{3}(\lambda)$). Also, the residue of $\Lambda^{(m)}(\lambda)$ at $\lambda = - i$ should vanish, thus we obtain the following condition,
\be
A_{2}(-i) = A_{3}(-i)\,.
\ee
  We put all the above requirements together and we find that
\be
A_{1}(\lambda) =  \prod_{j=1}^{m_{1}} {\sinh \mu (\lambda + \lambda_{j}^{(1)}) \over \sinh \mu(\lambda + \lambda_{j}^{(1)} + i)}\ {\sinh \mu (\lambda - \lambda_{j}^{(1)} - i) \over \sinh \mu(\lambda - \lambda_{j}^{(1)})}\,,
\ee
\be
A_{2}(\lambda) &=&  \prod_{j=1}^{m_{1}} {\sinh \mu (\lambda + \lambda_{j}^{(1)}+2i) \over \sinh \mu(\lambda + \lambda_{j}^{(1)} + i)}\ {\sinh \mu (\lambda - \lambda_{j}^{(1)} + i) \over \sinh \mu(\lambda - \lambda_{j}^{(1)})}\, \non\\ 
& & \prod_{j=1}^{m_{2}} {\sinh \mu (\lambda + \lambda_{j}^{(2)}+i) \over \sinh \mu(\lambda + \lambda_{j}^{(2)} + 2i)}\ {\sinh \mu (\lambda - \lambda_{j}^{(2)} - i) \over \sinh \mu(\lambda - \lambda_{j}^{(2)})}\,, 
\ee
\be
A_{3}(\lambda) =  \prod_{j=1}^{m_{2}} {\sinh \mu (\lambda + \lambda_{j}^{(2)}+ 3i) \over \sinh \mu(\lambda + \lambda_{j}^{(2)} + 2i)}\ {\sinh \mu (\lambda - \lambda_{j}^{(2)} + i) \over \sinh \mu(\lambda - \lambda_{j}^{(2)})}\,.
\ee
We obtain  $\bar A_{i}(\lambda)$ from (\ref{crosp}). It is easy to check that the eigenvalues satisfy all the above conditions.
Moreover we want the eigenvalues to be analytical, so the poles must vanish. This condition leads to the Bethe ansatz equations
\be
e_{1}(\lambda_{i}^{(1)})^{2N} &=& \prod_{i \ne j=1}^{m_{1}} e_{2}(\lambda_{i}^{(1)} - \lambda_{j}^{(1)})\ e_{2}(\lambda_{i}^{(1)} + \lambda_{j}^{(1)})\ \prod_{j=1}^{m_{2}} e_{-1}(\lambda_{i}^{(1)} - \lambda_{j}^{(2)})\ e_{-1}(\lambda_{i}^{(1)} + \lambda_{j}^{(2)})\, \non\\
1 &=& \prod_{i \ne j=1}^{m_{2}}e_{2}(\lambda_{i}^{(2)} - \lambda_{j}^{(2)})\ e_{2}(\lambda_{i}^{(2)} + \lambda_{j}^{(2)})\ \prod_{j=1}^{m_{1}} e_{-1}(\lambda_{i}^{(2)} - \lambda_{j}^{(1)})\ e_{-1}(\lambda_{i}^{(2)} + \lambda_{j}^{(1)})\,
\label{BAE}
\ee
where we have defined $e_{n}(\lambda)$ as
\be
e_{n}(\lambda) = {\sinh \mu(\lambda + {in \over 2}) \over \sinh \mu(\lambda - {in \over 2})}\,.
\ee
 The exact computation for the general case becomes complicated, however, one can ``guess'' the form of the general eigenvalue having in mind all the conditions that it must satisfy. The expression for the spectrum of the transfer matrix for any $\n$ is given by
\be
\Lambda^{(m)}(\lambda)   \propto  b(\lambda)^{2N} \sum_{k=1}^{\n}{\sinh(2 \mu \lambda) \over \sinh \mu(2\lambda + (k-1) i)} {\sinh \mu (2\lambda + i) \over \sinh \mu(2\lambda +ki)} A_{k}(\lambda)\,,
\label{spectrum}
\ee 
where
\be
A_{k}(\lambda) & = & \prod_{j=1}^{m_{k-1}} {\sinh \mu (\lambda + \lambda_{j}^{(k-1)} + ki) \over \sinh \mu(\lambda + \lambda_{j}^{(k-1)} +(k-1) i)} {\sinh \mu (\lambda - \lambda_{j}^{(k-1)} + i) \over \sinh \mu(\lambda - \lambda_{j}^{(k-1)})} \non\\ 
& & \prod_{j=1}^{m_{k}} {\sinh \mu (\lambda + \lambda_{j}^{(k)} + (k-1)i) \over \sinh \mu(\lambda + \lambda_{j}^{(k)} + ki)} {\sinh \mu (\lambda - \lambda_{j}^{(k)} - i) \over \sinh \mu(\lambda - \lambda_{j}^{(k)})}\,, \qquad k = 1, \ldots, \n\,,
\label{A}
\ee 
$m_{0} = N$ and $m_{\n} = 0$. Also, $\bar \Lambda^{(m)}(\lambda) = \Lambda^{(m)}(-\lambda - \rho)$. The procedure we described uniquely fixes the ``dressing'' functions. By inspection we can verify that the above eigenvalues indeed satisfy all the requirements we derived previously i.e, analyticity, asymptotic behaviour, crossing, etc. e.g.,
\be
\bar A_{k}(\lambda) = A_{k}(-\lambda - \rho)\,, \qquad k=1, \ldots, \n\,,
\ee
\be
A_{1}(\lambda) A_{1}(-\lambda) = 1\,, \qquad \bar A_{\n}(\lambda) \bar A_{\n}(-\lambda) = 1\,.
\ee
 We can see from (\ref{A}) that every two terms have the same poles. From the analyticity of the eigenvalues we obtain the Bethe ansatz equations 
\be
1 &=& \prod_{i \ne j=1}^{m_{k}}e_{2}(\lambda_{i}^{(k)} - \lambda_{j}^{(k)})\ e_{2}(\lambda_{i}^{(k)} + \lambda_{j}^{(k)}) \non\\
& & \prod_{j=1}^{m_{k+1}} e_{-1}(\lambda_{i}^{(k)} - \lambda_{j}^{(k+1)})\ e_{-1}(\lambda_{i}^{(k)} + \lambda_{j}^{(k+1)}) \non\\
& &\prod_{j=1}^{m_{k-1}} e_{-1}(\lambda_{i}^{(k)} - \lambda_{j}^{(k-1)})\ e_{-1}(\lambda_{i}^{(k)} + \lambda_{j}^{(k-1)})\,, \non\\
& & k = 1, \ldots, \n\,,
\label{BAEP}
\ee
for $\n = 3$ we recover (\ref{BAE}).
The results, as expected, coincide with the known ones obtained by nesting
\cite{DVGR3}, (\cite{DVGR2} for $\xi \rightarrow \pm i\infty$). 

\section{Discussion}
We generalised the fusion procedure for open spin chains without crossing symmetry. Furthermore, we showed that even though the $R$ matrix does not have crossing symmetry the transfer matrix satisfies a crossing property (\ref{cross}). We applied these results to diagonalise the transfer matrix via the analytical Bethe ansatz method. We found explicit expressions for the transfer matrix spectrum (\ref{spectrum}) and we deduced the Bethe ansatz equations (\ref{BAEP}), avoiding nesting. The main realization in this paper was the necessity of the transfer matrix $\bar t(\lambda)$ (\ref{transfer12}) in the derivation of the analytical Bethe ansatz. Indeed, it was necessary to consider $\bar t(\lambda)$ together with the usual transfer matrix in order to derive the conditions that the eigenvalues should satisfy. 

Here, we considered the special case where the chain has a $U_{q} \left(SU(\n) \right)$ symmetry. However, we believe that the previous analysis can be extended even in the case of the reduced symmetry $U_{q}\left(SU(l)\right) \times U_{q}\left(SU(\n - l)\right) \times U(1)$ \cite{dn/duality}. Moreover, the Bethe ansatz equations are known for open spin chains with ``soliton preserving'' boundary conditions. There is also the case of ``soliton non-preserving'' boundary conditions (see e.g., \cite{sklyanin}, \cite{delius1}) for which the Bethe ansatz equations are not known. Using the analytical Bethe ansatz, one can presumably derive the corresponding transfer matrix spectrum and the Bethe ansatz equations, avoiding nesting. We hope to address these questions in a future work \cite{new}.

\section{Acknowledgements}
I am grateful to P. Bowcock, E. Corrigan and G.W. Delius for helpful discussions. I would also like to thank R.I. Nepomechie for valuable suggestions and for prior collaborations. This work was supported by the European Commission under the TMR Network ``Integrability, non-perturbative effects, and symmetry in quantum field theory'', contract number FMRX-CT96-0012.

\end{document}